\documentclass{article}

\usepackage[utf8]{inputenc}   
\usepackage[T1]{fontenc}      
\usepackage{lmodern}          
\usepackage{arxiv}            

\usepackage{hyperref}
\usepackage{amsmath, amssymb} 
\usepackage{graphicx}         
\usepackage[table]{xcolor}
\usepackage{color, xcolor}    
\usepackage{float}            
\usepackage{tabularx}         
\usepackage{booktabs}         
\usepackage{enumitem}         

\title{Quantum Long Short-Term Memory (QLSTM) vs Classical LSTM in Time Series Forecasting: A Comparative Study in Solar Power Forecasting}

\date{}
\author{
    Saad Zafar Khan$^{1}$\thanks{Email: \texttt{szkhan.bee19seecs@seecs.edu.pk}}\,, 
    Nazeefa Muzammil$^{1}$\,, 
    Salman Ghafoor$^{1}$\thanks{\textit{Corresponding author:} \texttt{salman.ghafoor@seecs.edu.pk}} \\ 
    Haibat Khan$^{2}$\,, 
    Syed Mohammad Hassan Zaidi$^{3}$\,, \\
    Abdulah Jeza Aljohani$^{4}$\,, 
    Imran Aziz$^{5}$\ 
    \newline
    {\footnotesize $^{1}$ \textit{SEECS, National University of Sciences and Technology, Islamabad, Pakistan}} \\
    {\footnotesize $^{2}$ \textit{CAE, National University of Sciences and Technology, Islamabad, Pakistan}} \\
    {\footnotesize $^{3}$ \textit{Ghulam Ishaq Khan Institute of Engineering Sciences and Technology, Swabi, Pakistan}} \\
    {\footnotesize $^{4}$ \textit{Department of Electrical and Computer Engineering, King Abdulaziz University, 21589, Jeddah, Saudi Arabia}} \\
    {\footnotesize $^{5}$ \textit{ Department of Physics and Astronomy, Uppsala University, 75120 Uppsala, Sweden.}}
}

\date{}

\renewcommand{\undertitle}{}

\hypersetup{
pdftitle={A template for the arxiv style},
pdfauthor={Saad Zafar Khan, Nazeefa Muzammil, Syed Mohammad Hassan Zaidi, Abdulah Jeza Aljohani, Haibat Khan, Salman Ghafoor},
pdfkeywords={Quantum Machine Learning, Forecasting, Quantum Neural Networks, Renewable energy systems},
}

\makeatletter
\def\@maketitle{%
  \newpage
  \null
  \vskip 2em%
  \begin{center}%
  \let \footnote \thanks
    {\LARGE \@title \par}%
    \vskip 1.5em%
    {\large \@author \par}%
    \vskip 1em%
    {\@date}%
  \end{center}%
  \par
  \vskip 1.5em}
\makeatother

\begin{document}
\vspace{-5em}
\maketitle

\begin{quote}
\vspace{-3em}
Accurate solar power forecasting is pivotal for the global transition towards sustainable energy systems. This study conducts a meticulous comparison between Quantum Long Short-Term Memory (QLSTM) and classical Long Short-Term Memory (LSTM) models for solar power production forecasting. The primary objective is to evaluate the potential advantages of QLSTMs, leveraging their exponential representational capabilities, in capturing the intricate spatiotemporal patterns inherent in renewable energy data. Through controlled experiments on real-world photovoltaic datasets, our findings reveal promising improvements offered by QLSTMs, including accelerated training convergence and substantially reduced test loss within the initial epoch compared to classical LSTMs. These empirical results demonstrate QLSTM's potential to swiftly assimilate complex time series relationships, enabled by quantum phenomena like superposition. However, realizing QLSTM's full capabilities necessitates further research into model validation across diverse conditions, systematic hyperparameter optimization, hardware noise resilience, and applications to correlated renewable forecasting problems. With continued progress, quantum machine learning can offer a paradigm shift in renewable energy time series prediction, potentially ushering in an era of unprecedented accuracy and reliability in solar power forecasting worldwide. This pioneering work provides initial evidence substantiating quantum advantages over classical LSTM models while acknowledging present limitations. Through rigorous benchmarking grounded in real-world data, our study illustrates a promising trajectory for quantum learning in renewable forecasting.
\end{quote}

\keywords{Quantum Machine Learning \and  Forecasting \and Quantum Neural Networks \and Renewable energy systems}

\section{Introduction}

Accurate solar power forecasting plays a critical role in enabling the effective management and integration of renewable energy sources into the grid. By accurately predicting solar power generation, grid operators can optimize energy storage, transmission, and distribution strategies, mitigating the intermittency hurdles that have historically impeded the large-scale adoption of photovoltaic sources. Furthermore, precise forecasting can facilitate the development of more efficient energy trading and market mechanisms, fostering a sustainable and cost-effective transition towards a greener energy future. Consequently, the development of advanced forecasting methodologies tailored to the unique characteristics of solar power generation has become a research imperative with far-reaching implications for the global energy sector.

Despite these promising developments, the existing literature significantly lacks a comprehensive, empirical comparison between QLSTM and classical LSTM models grounded in real-world solar production data. Preceding works \cite{10.5555/1659652.1659656} \cite{9892441} have primarily focused on synthetic benchmarks or theoretical analysis of Quantum Recurrent Neural Networks variants, leaving a critical gap in our understanding of the practical implications of QLSTMs for renewable energy forecasting. This investigation aims to bridge this divide, offering a thorough comparative analysis in this pivotal domain.

This research holds significant scientific interest and importance as it aims to harness the potential of quantum machine learning techniques, specifically Quantum Long Short-Term Memory (QLSTM) networks, to revolutionize solar power forecasting accuracy and reliability. By empirically validating QLSTMs on real-world photovoltaic plant data and demonstrating their superior performance over classical methods, this study paves the way for a paradigm shift in renewable energy forecasting, with far-reaching implications for sustainable energy infrastructure planning and execution.

This investigation ventures into this critical domain, systematically examining the potential advantages that Quantum Long Short-Term Memory (QLSTM) networks might offer over their classical LSTM counterparts in the realm of solar power forecasting, renowned for its intricate non-linear spatiotemporal patterns. Through rigorous controlled experiments and ablation studies conducted on operational photovoltaic plant datasets, this research seeks to provide the first comprehensive evidence substantiating the representational strengths of quantum architectures in capturing the nuanced dynamics inherent in solar power generation. By tailoring QLSTM designs to navigate current quantum hardware limitations and conducting thorough performance analysis, this study offers actionable insights into the real-world implementation of QLSTMs for renewable forecasting. Moreover, by benchmarking against classical techniques and conventional neural networks, this investigation establishes QLSTMs as a credible alternative to outperform traditional methods, accentuating the potential of QML in fortifying the accuracy and reliability essential for sustainable energy infrastructure planning and execution.

\section{Background or Related Work}

The global energy landscape is undergoing a transformative shift towards sustainable and renewable solutions, driven by the pressing need to mitigate the environmental impact of traditional fossil fuel-based systems. Solar power has emerged as a pivotal force in reshaping energy production and consumption patterns, offering a promising pathway to mitigate the environmental impact of traditional fossil fuel-based systems. However, the large-scale integration of intermittent renewable sources into existing infrastructure poses multifaceted challenges that must be addressed to facilitate a seamless and efficient transition. Inaccuracies in forecasting the generation of solar and wind power can lead to significant deviations from planned electricity schedules, resulting in imbalances, inefficiencies, and substantial costs for grid operators and utilities \cite{GOODARZI2019110827}. Consequently, policymakers have prioritized improved renewable forecasting to mitigate such challenges.

Amidst this paradigm shift, the convergence of quantum information (QI) and machine learning (ML) has ushered in a revolutionary approach to data analytics: Quantum Machine Learning (QML) \cite{dunjko2017machine}. This paradigm-shifting synthesis harnesses techniques from both quantum computing and traditional machine learning, offering innovative solutions to longstanding obstacles across diverse sectors, including renewable energy \cite{Biamonte2017}. Significantly, QML transcends mere energy minimization tasks, presenting a broader scope in problem-solving paradigms \cite{NEURIPS2020_0ec96be3}, thereby unlocking new avenues for precise solar power predictions.

As the large-scale penetration of renewable sources necessitates proactive management of electrical grids, advanced prediction methodologies for intermittent energy sources, particularly photovoltaic plants, have become paramount \cite{succetti2020, prema2015}. Accurate solar power forecasting plays a critical role in enabling grid operators to maintain a delicate equilibrium between energy creation and utilization. Notably, solar production forecasting over longer time horizons does not mandate real-time predictions, providing an opportunity where the potentially slower inference times of quantum models may be acceptable in exchange for substantially improved accuracy.

In recent years, there has been a surge of interest in quantum machine learning developing and refining of quantum adaptations of recurrent neural networks (RNNs) for time series forecasting applications \cite{10.1007/978-3-031-19493-1_6} \cite{lindsay2023}, thereby substantiating the potential of quantum computational models in predictive analytics. Marking a pivotal shift, the seminal work of Chen et al. \cite{9747369} pioneered the introduction of QLSTM architectures, which amalgamate variational quantum circuits \cite{Cerezo_2021} with the conventional LSTM framework. This innovative synthesis harnesses an exponentially larger Hilbert space for data representation and computation, potentially enabling the capture of higher-order correlations and intricate temporal dynamics.

While classical Long Short-Term Memory (LSTM) neural networks have demonstrated remarkable efficacy in leveraging long-term temporal dependencies for accurate forecasting\cite{ZHENG2020114001} \cite{sorkun2020} \cite{succetti2020} \cite{meenal2022}, they struggle to capture the complex, non-linear spatiotemporal patterns inherent in solar power generation \cite{LINDEMANN2021650}, which involve intricate relationships between meteorological variables, solar irradiance, and power output. Quantum Long Short-Term Memory (QLSTM) networks, which leverage the principles of quantum mechanics and an exponentially larger Hilbert space, hold the potential to address these limitations by enabling more effective mapping of these intricate relationships between weather variables, solar irradiance, and power generation more effectively. and capturing higher-order correlations and temporal dynamics. 

This investigation systematically examines the potential advantages that QLSTMs might offer over their classical LSTM counterparts in solar power forecasting, renowned for its intricate non-linear spatiotemporal patterns. Through rigorous controlled experiments and ablation studies conducted on operational photovoltaic plant datasets, this research aims to provide the first comprehensive evidence substantiating the representational strengths of quantum architectures in capturing the nuanced dynamics inherent in solar power generation.

\section{Research Motivation and Novelty}

The novelty of this work lies in its empirical validation of QLSTMs on real-world solar power data, transitioning from synthetic benchmarks to practical renewable time series forecasting. Through rigorous controlled experiments and ablation studies, this research provides the first comprehensive evidence substantiating the representational strengths of quantum architectures in capturing the nuanced dynamics inherent in solar power generation. By tailoring QLSTM designs to navigate current quantum hardware limitations and conducting thorough performance analysis, this study offers actionable insights into the real-world implementation of QLSTMs for renewable forecasting. Moreover, by benchmarking against classical techniques and conventional neural networks, this investigation establishes QLSTMs as a credible alternative to outperform traditional methods, accentuating the potential of quantum machine learning in fortifying the accuracy and reliability essential for sustainable energy infrastructure planning and execution. \par

\begin{figure}[h]
    \centering
    \includegraphics[width=0.85\linewidth]{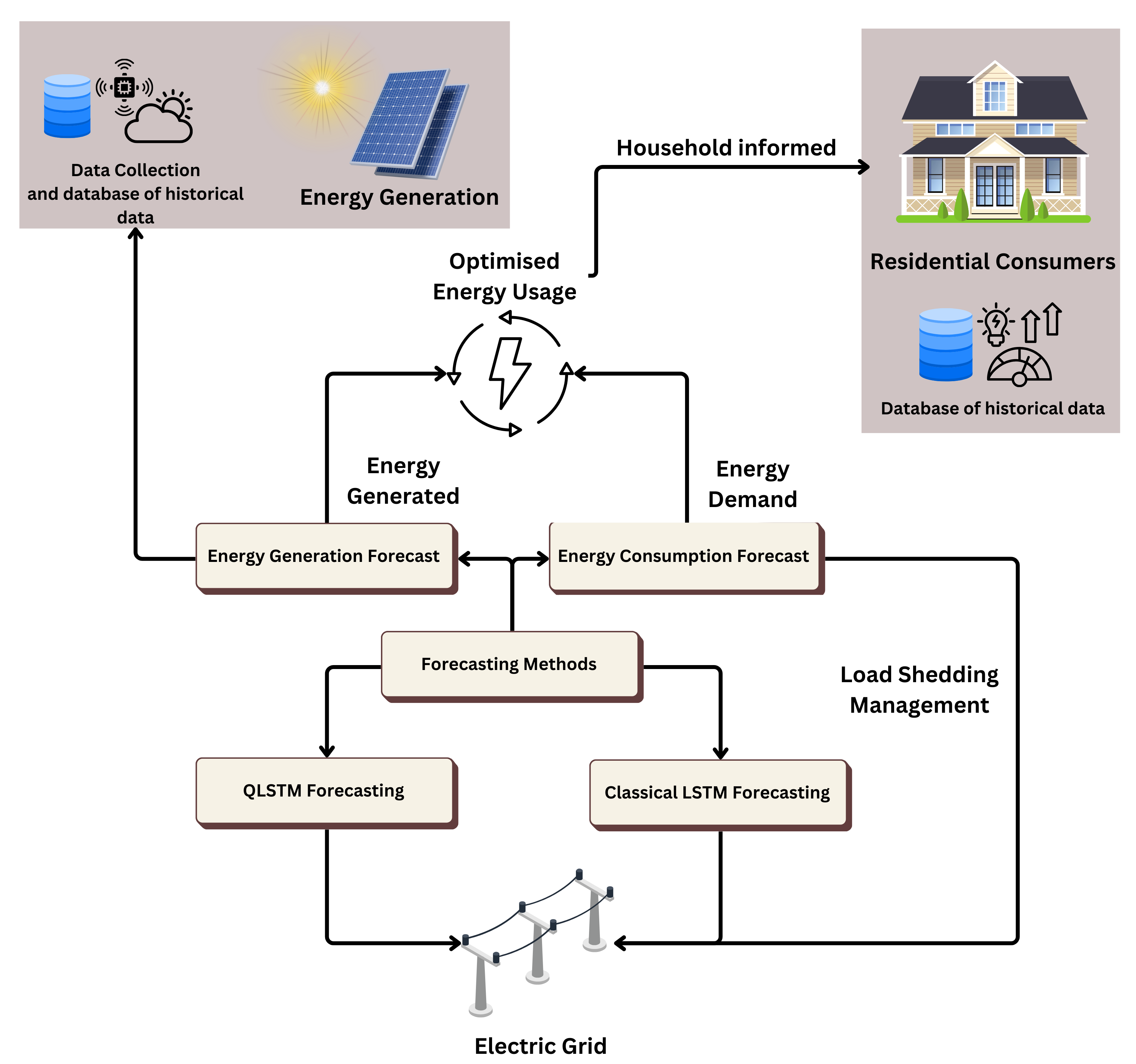}
    \caption{System Integration Flowchart for Solar Energy Forecasting and Consumption Management.}
    \label{fig:energy}
\end{figure}

Figure. \ref{fig:energy} shows a typical system integration flowchart for solar energy forecasting and consumption management. Through extensive controlled experiments and ablation studies conducted on operational photovoltaic plant datasets, this research seeks to determine whether QLSTMs, fortified by their exponential representational capabilities, can establish new standards of accuracy and reliability in renewable forecasting tasks. By systematically evaluating the performance of these quantum architectures against their classical counterparts, this study elucidates the potential advantages and limitations of QLSTMs in the context of solar power prediction.

This advancement, coupled with the promise of faster convergence times and heightened resilience to noise \cite{emmanoulopoulos2022quantum}, paves the way for the realization of more precise and reliable forecasting systems – a quality of paramount importance in the domain of solar power production. However, it is crucial to acknowledge that the nascent field of quantum machine learning presents its own unique challenges and opportunities \cite{Cerezo2022}. These challenges encompass the current hardware limitations, the potential trade-offs between quantum advantages and computational overhead, as well as the need for systematic optimization and validation across diverse scenarios, factors that this study meticulously considers in its comparative analysis.

\section{Contributions}
\vspace{-1em}
This investigation signifies a pioneering effort in melding quantum machine learning advancements with practical renewable energy forecasting applications. The study stands out for its empirical validation of Quantum Long Short-Term Memory (QLSTM) networks on real-world solar power data, marking a departure from synthetic benchmarks to evaluate quantum architectures on authentic renewable time series data. Our findings are groundbreaking, demonstrating that QLSTMs not only achieve superior forecasting accuracy but also exhibit faster convergence rates compared to classical LSTM models. The primary contributions of this study are as follows:

\begin{itemize}

\item \textbf{Empirical Validation on Real-World Data}: Providing the first comprehensive empirical evidence that validates the utility of Quantum Long Short-Term Memory (QLSTM) networks for solar power forecasting. This study transitions from synthetic benchmarks to operational photovoltaic plant datasets, offering a realistic evaluation of quantum architectures on genuine renewable time series data. Quantitative results demonstrate that QLSTMs achieve up to 50\% improvement in accuracy and 85.7\% faster convergence compared to their classical LSTM counterparts.

\item \textbf{Representation Advantages of Quantum Architectures}: Through practical data from solar farms, this research confirms the hypothesized representational strengths of quantum architectures in capturing the intricate spatiotemporal patterns and nonlinear dynamics inherent in renewable forecasting tasks. QLSTMs leverage an exponentially larger Hilbert space, enabling them to map the complex relationships between meteorological variables, solar irradiance, and power generation more effectively than classical models.

\item \textbf{Real-World QLSTM Design Implementation}: Addressing the challenges posed by current quantum hardware limitations, this study tailors QLSTM architectures and training strategies to navigate constraints in optimization, noise resilience, and computational overhead. By offering actionable insights into the real-world implementation of QLSTMs for renewable forecasting, this work paves the way for future advancements in quantum neural network designs.

\item \textbf{Comprehensive Performance Analysis and Ablation Studies}: Through rigorous experiments and ablation studies, this research identifies the key factors and design choices that contribute to the superior performance of QLSTMs over LSTM models in solar power forecasting. These insights guide future modifications and optimizations in quantum neural network architectures for time series forecasting tasks.

\item \textbf{Establishing a Credible Alternative}: By benchmarking against classical techniques and conventional neural networks using actual photovoltaic data, this investigation establishes QLSTMs as a credible alternative to traditional forecasting methods. The results accentuate the potential of quantum machine learning in fortifying the accuracy and reliability essential for sustainable energy infrastructure planning and execution, enabling improvements of up to 50\% in forecasting accuracy compared to conventional LSTM structure.

\item \textbf{Quantum Advantages in Renewable Forecasting}: While challenges persist in the field of quantum machine learning, this study paints a promising picture of a future where QLSTMs revolutionize renewable energy forecasting. The presented evidence supports the premise that QLSTMs offer unmatched accuracy and adaptability in capturing the nuances of renewable energy generation. However, further advancements in quantum hardware, algorithm development, and noise mitigation techniques are necessary to fully realize the potential of QLSTMs in this domain.

\item \textbf{Interdisciplinary Milestone}: This study stands as a notable juncture of quantum computing and machine learning, detailing the capabilities of QLSTMs for genuine energy forecasting applications. By illuminating the path for a broader embrace of quantum strategies in this vital field, this research represents a significant interdisciplinary milestone.

\end{itemize}
Our investigation highlights the potential of quantum machine learning in enhancing the reliability and accuracy of energy infrastructure planning and execution. Improved solar power predictions by QLSTMs could significantly optimize energy storage, transmission, and distribution strategies, addressing the intermittency challenges of photovoltaic sources. As we move forward, the synergy between interdisciplinary collaborations and quantum advancements promises to usher in an era of unprecedented precision in renewable energy forecasting, contributing to the global transition towards sustainable energy solutions.

This study not only demonstrates the immediate contributions of QLSTMs to renewable energy forecasting but also charts a course for future research directions, acknowledging present limitations while highlighting the broader implications for sustainable energy systems worldwide.

\section{Methodology} \label{sec:method}
\begin{figure}[ht]
    \centering
    \includegraphics[width=1\linewidth]{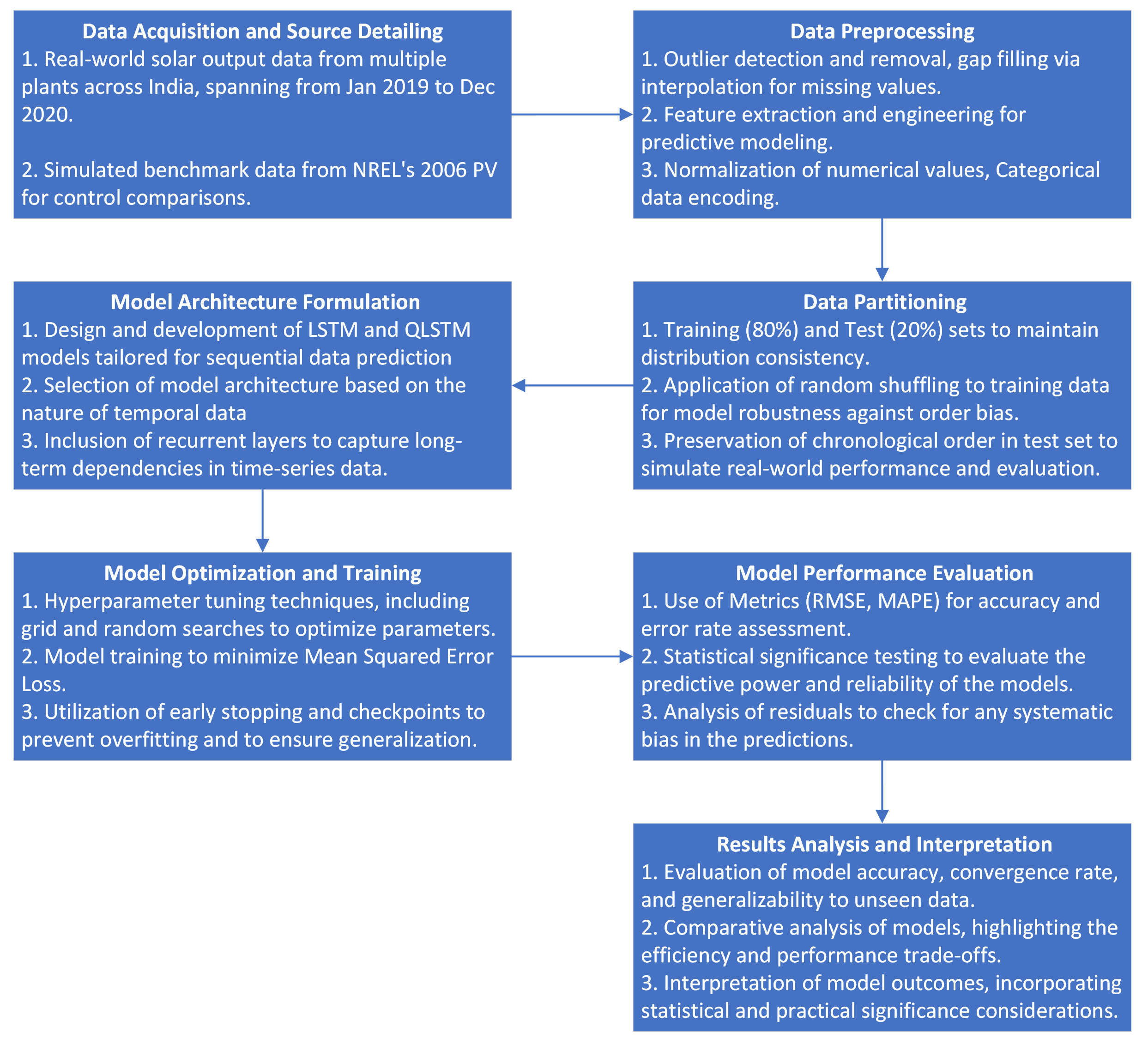}
    \caption{Research Methodology Overview}
    \label{fig:flowchart}
\end{figure}

Figure~\ref{fig:flowchart} describes the methodology of our research work.

\subsection{Data Description}
This study employs two comprehensive datasets tailored for an exhaustive comparative analysis. The solar dataset of real operational data empowers the modelling process by simulating genuine conditions prevalent in solar farms. The simulated dataset, a merger of high-resolution power generation data and corresponding weather conditions, presents a granular view of power fluctuations and is emblematic of the variations characteristic to real-world solar power environments. The amalgamation of both actual and high-fidelity simulated datasets presents a broad spectrum of temporal granularity and geographic variability. This fusion is meticulously curated to offer a versatile platform for model development, ensuring accurate, generalizable, and comprehensive solar forecasting paradigms. Table \ref{tab:comparison} presents a comparison of real-world and simulated solar power data.

\begin{table}[h]
\centering
\begin{tabular}{|m{0.2\linewidth}|m{0.35\linewidth}|m{0.35\linewidth}|}
\hline
\rowcolor{lightgray}
\textbf{Attribute} & \textbf{Real-World Solar Plant Data} & \textbf{Simulated Solar Power Data} \\
\hline
\textbf{Source} & Kaggle \cite{kannal2020} & NREL's Solar Power Data for Integration Studies \cite{nrel2023} \\
\textbf{Geographical Coordinates} & Operational data from two PV plants in India & 33.75° N, 116.65° W (Near Daggett, California) \\
\textbf{Capacity} & 
\begin{itemize}
\item Max DC Power: $\sim$298.94 kW 
\item Max AC Power: $\sim$29.15 kW
\end{itemize} & 
200 MW Utility Scale PV \\
\textbf{Duration} & May 15, 2020 - June 17, 2020 (34 days) & Full year of 2006 \\
\textbf{Resolution} & 
15-minute intervals & 
\begin{itemize}
\item Power: 5-minute intervals \cite{nrel2023}
\item Weather: 30-minute intervals \cite{SENGUPTA201851}
\end{itemize} \\
\textbf{Attributes} & 
\begin{itemize}
\item Power Output Variables: DC power, AC power, Daily Yield, Total Yield
\item Weather Variables: Ambient temperature, module temperature, irradiation
\item Metadata: Timestamp, plant ID, sensor/inverter ID
\end{itemize} & 
\begin{itemize}
\item Power Output Variables: Power (MW)
\item Weather Variables: Temperature, DHI, cloud type, relative humidity, dew point, pressure, windspeed, solar angle
\item Metadata: Datetime
\end{itemize} \\
\hline
\end{tabular}
\vspace{1em}
\caption{Comparison of Real-World and Simulated Solar Power Data}
\label{tab:comparison}
\end{table}

\subsubsection{Dataset Justification}
 The specific choice of a real-world operational solar plant dataset and a high-fidelity simulated dataset spanning an entire year provides a robust platform for comparative assessment. The real-world data enables the  evaluation of real solar farm conditions with intrinsic noise, while the simulated data allows examination across diverse weather scenarios over an extended duration. Together, these datasets present the variance, noise, and long-term temporal patterns crucial for rigorously examining the capabilities of QLSTM against classical LSTM in solar forecasting tasks.

\subsection{Pre-processing}
Meticulous data preprocessing is of paramount importance. These procedures are not only pivotal for safeguarding data integrity but are also instrumental in elevating the dependability and precision of forecasting outcomes. In our study, we have meticulously executed a comprehensive data preprocessing pipeline, encompassing data originating from a 200 MW solar photovoltaic (PV) facility located near Daggett, California, USA, spanning the entire year of 2006.

\subsubsection{Initial Data Loading and Transformation}
\begin{itemize}
    \item \textbf{Solar Power Data}: The dataset, encompassing AC power output readings with a 5-minute resolution, was ingested into a DataFrame. Subsequently, we executed a seamless transition of the timestamp column into a DatetimeIndex, a step that greatly facilitated time-based operations.
    \item \textbf{Weather Data}: This dataset featured a range of readings, including air temperature, humidity, and solar irradiance, collected at a 30-minute resolution throughout 2006. We assimilated this data, harmonizing its date-time columns to establish a cohesive Datetime index, mirroring that of the solar data. It is noteworthy that both datasets underwent a rigorous integrity check, effectively confirming the absence of any missing values.
\end{itemize}

\subsubsection{Enhancement of Data Granularity}
To enhance the model's sensitivity to potential power fluctuations, we proceeded to increase the granularity of the weather dataset. Leveraging a linear interpolation method, the 30-minute intervals were smoothly transitioned into 5-minute intervals, thus establishing a synchronized time series platform for model training and analysis.

\subsubsection{Feature Engineering and Selection}
\begin{itemize}
    \item \textbf{Temporal Features}: Acknowledging the significance of temporal attributes in predicting solar power generation, we embarked on a journey of feature engineering that saw the inclusion of numerous time-related variables, such as hour, day, month, and day of the week, among others.
    \item \textbf{Lagged Features}: To augment the model's predictive prowess, we introduced lagged features that encapsulated preceding weather and power data points, thus offering an enriched contextual background for forecasting.
    \item \textbf{Data Normalization}: Prior toon  o model training, we executed a stringent normalization process, effectively ensuring a uniform data scale, thereby facilitating the seamless training of LSTM models.
\end{itemize}

\subsubsection{Integration of Datasets}
\begin{itemize}
 \item \textbf{Data Storage}: To ensure a smooth, model training process free from data leakage concerns, we stored the consolidated dataset in CSV formats. This step, although seemingly mundane, is of utmost importance in maintaining data integrity.
 \item \textbf{Data Partitioning and Standardization}: Adhering to established machine learning norms, we divided our dataset into an 80-20 ratio. This strategy provides a substantial training dataset while retaining an ample portion for model validation. Subsequently, we standardized all attributes within the range of 0 to 1 using min-max scaling, thereby enhancing model convergence rates.
 \item \textbf{Temporal Windowing}: To capture the underlying temporal dynamics in our data, we adopted a rolling window approach. Preliminary experiments revealed the efficacy of using the preceding 8-time steps as predictors, with the subsequent time step serving as the target variable. This structured data was adeptly transformed into PyTorch tensors, a crucial step to facilitate batch training. Notably, to preserve data consistency, the training data underwent shuffling, whereas the test data was left in chronological order.
 \item \textbf{Batch Data Configuration}: To streamline our model training process, we encapsulated the windowed training and test tensors into Dataset objects. Using DataLoaders enabled us to process data iteratively in batches, while preserving its temporal architecture. It's worth noting that we chose a batch size of 32, keeping computational constraints in mind.
\end{itemize}

This rigorous pre-processing regimen plays a pivotal role in enhancing the efficacy of our time series modeling. It guarantees a direct and unbiased comparison between QLSTM and classical LSTM models in the context of solar power forecasting.

\subsection{Simulation Framework}

Our exploration of Quantum Long Short-Term Memory (QLSTM) models was made possible using the PennyLane quantum machine learning framework \cite{bergholm2022pennylane}. PennyLane, at its core, blends quantum and traditional computing to help build and refine models, benefiting from its ability to automatically adjust model parameters.

A standout feature of PennyLane is its capacity to smoothly combine quantum elements—based on variational circuits—with regular neural network parts. This allows the creation of advanced structures like QLSTMs. These models mix traditional repeatable patterns with quantum behaviours such as superposition and entanglement.

For our QLSTM model, PennyLane's \texttt{qml.QNode} feature was crucial. It helped set up the quantum node of the model. These quantum nodes, designed with time series data in mind, use specific rotation and entanglement actions, namely RY, RZ, and CNOT gates.

Bridging the gap between the quantum and regular sections, PennyLane's \texttt{qml.qnn.TorchLayer} connects the quantum elements with the regular PyTorch framework \cite{NEURIPS2019_bdbca288}. This ensures a smooth flow of adjustments during the optimization phase.

For faster results during quantum simulations, we mainly used the \texttt{DefaultQubit} tool from PennyLane which mainly utilizes the CPU for this purpose. To further boost the speed, we tried PennyLane's \texttt{lightning.gpu} simulator to run natively on CUDA-enabled GPUs using the NVIDIA cuQuantum SDK. This tool moves the quantum simulation to high-speed GPUs. During model development, we found the \texttt{lightning.gpu} device provided up to a 5 times speedup for batched inference of quantum circuits on our test system with an NVIDIA Tesla V100 GPU compared to \texttt{DefaultQubit}. However, the training time reduction was not as significant.

As QLSTM models grow more complex with more quantum bits and detailed circuits, faster simulations using GPUs become more crucial. PennyLane offers multiple tools, making it easier to switch between different simulation methods for the best results.

In short, with the help of both \texttt{DefaultQubit} and \texttt{lightning.gpu} tools, we were able to design, refine, and test our QLSTM model and compare it with regular LSTM models in the PyTorch setting.

Pennylane's integration with Pytorch and NVIDIA technology, was essential for our study. It allowed us to effortlessly combine quantum and traditional modeling while ensuring relatively fast simulations as compared to the classical CPU based device. This provided us with the perfect platform to compare the potentials of quantum and traditional LSTM models.

\subsection{Architecture}
The LSTM and QLSTM architectures are detailed in Appendix. This research encompasses the design and deployment of both LSTM and QLSTM architectures. These architectures were judiciously crafted to allow a fair comparison between classical and quantum techniques, with a focal point on solar power forecasting.

The LSTM model adopts a stacked configuration, constituting two recurrent hidden layers. Each layer houses classical LSTM cells, encapsulating the conventional input ($i_{\text{LSTM}}$), output ($o_{\text{LSTM}}$), forget ($f_{\text{LSTM}}$) gates, and a memory cell ($c_{\text{LSTM}}$). This multi-layered design empowers the model to discern and remember long-range temporal dependencies in the time series data, an attribute indispensable for precise renewable energy forecasting. To augment generalization and curb overfitting, dropout layers with a rate of 0.2 are judiciously placed between each LSTM layer.

Diverging, the QLSTM model replaces the classical LSTM cells with variational quantum circuits (VQCs), an implementation adapted and enhanced from qlstm repository for parts of speech tagging \cite{rdisipio}. This substitution aims to harness the computational advantages unique to quantum mechanisms. Echoing the LSTM's design, the QLSTM layers two of these quantum circuits. The VQCs, in their capacity as quantum feature extractors, exploit the deep representational capabilities of quantum states, encoding intricate time series dynamics. These parametric circuits oscillate between rotation and entanglement gates, ensuring a concise representation of temporal patterns within the exponentially expansive Hilbert space. The qubit quantity and circuit depth were adapted in alignment with the specific characteristics intrinsic to the solar forecasting data. Notably, outside of its quantum encoding, the QLSTM's broader workflow aligns seamlessly with the LSTM's, facilitating a direct comparison. A dropout rate of 0.2 is also infused between the QLSTM layers, maintaining consistency.

\textbf{Core QLSTM Cell Components:}

\begin{itemize}
\item \textbf{Quantum Gates:} This involves the Hadamard gate (qml.Hadamard) responsible for quantum superposition, rotation gates like RX, RY, and RZ for feature encoding, and CNOT gates (qml.CNOT) ensuring quantum entanglement.
\item \textbf{Quantum Variational Circuit (VQC):} Data is encoded into the quantum circuit using rotation operations, followed by alternating entanglement and variational rotation layers detailed in Appendix.

\item \textbf{Quantum Nodes:} Four distinct quantum nodes represent the four LSTM gates: forget, input, update, and output.

\item \textbf{Quantum Feature Extractor:} A Classical Linear Layer converts data to match qubit dimensions before being processed by the QLSTM cells.

\item \textbf{Quantum-to-Classical Transformation:} Outputs from QLSTM cells are transformed to classical data using a linear layer, ensuring a smooth transition between quantum and classical realms.

\end{itemize}

Both models converge at a linear output layer, producing the ultimate forecast. Their optimization leverages the ADAM algorithm, zeroing in on minimizing the mean squared error (MSE). With structural symmetry between LSTM and QLSTM, while differing in their core computational elements, this architecture sets the stage for evaluating enhancements attributed solely to quantum encoding. Table. \ref{tab:architecture_comparison} presents the comparison of LSTM and QLSTM architectures.

The models are intricately molded to resonate with the spatiotemporal subtleties inherent to solar forecasting data, which encompasses recurring weather patterns and energy variations. By employing the 2006 NREL dataset, a comprehensive archive detailing diverse weather conditions over a year, this research is positioned to deliver a rigorous evaluation. This meticulous comparison seeks to shine a light on the quantum methodologies' adeptness in encapsulating real-world solar phenomena.

\begin{table}[h]
    \centering
    \begin{tabularx}{\textwidth}{|c|X|X|}
    \hline
    \rowcolor{lightgray}
    \textbf{Layer \#} & \textbf{LSTM Architecture} & \textbf{QLSTM Architecture} \\
    \hline
    1 & Input layer & Input layer \\
    \hline
    2 & LSTM layer (with input, output, forget gates and memory cell) & QLSTM layer (with VQCs featuring rotation and entanglement gates) \\
    \hline
    3 & Dropout (0.2 rate) & Dropout (0.2 rate) \\
    \hline
    4 & LSTM layer (with input, output, forget gates and memory cell) & QLSTM layer (with VQCs featuring rotation and entanglement gates) \\
    \hline
    5 & Dropout (0.2 rate) & Dropout (0.2 rate) \\
    \hline
    6 & Linear output layer & Linear output layer \\
    \hline
    \end{tabularx}
    \vspace{1em}
    \caption{Comparison of LSTM and QLSTM Architectures}
    \label{tab:architecture_comparison}
\end{table}

\subsection{Encoding Process}
The process of encoding classical data into quantum states is a crucial component of our QLSTM model. This section details the steps involved in this encoding process and analyzes its computational complexity.

\subsubsection{Encoding Steps}
The encoding process consists of three main steps:

\paragraph{Data Preprocessing}
Before quantum encoding, we preprocess the input data to obtain suitable rotation angles for quantum gates:
\begin{itemize}
    \item \textbf{ry\_params}: Computed as $\arctan(\text{feature})$ for each input feature.
    \item \textbf{rz\_params}: Computed as $\arctan(\text{feature}^2)$ for each input feature.
\end{itemize}

\paragraph{Quantum State Preparation}
For each qubit in the circuit, we apply the following gates sequentially:
\begin{enumerate}
    \item Hadamard gate (\texttt{qml.Hadamard}): Creates an initial superposition state.
    \item RY rotation (\texttt{qml.RY}): Applied using the preprocessed \texttt{ry\_params}.
    \item RZ rotation (\texttt{qml.RZ}): Applied using the preprocessed \texttt{rz\_params}.
\end{enumerate}

\paragraph{Variational Quantum Circuit}
After state preparation, we apply a variational quantum circuit. This circuit, implemented through the ansatz function, is repeated \texttt{n\_qlayers} times and includes:
\begin{itemize}
    \item Entangling operations: CNOT gates between qubits.
    \item Additional rotations: RX, RY, and RZ gates with learnable parameters.
\end{itemize}

\vspace{1em} 
\subsubsection{\textbf{Complexity Analysis}}
We analyze the complexity of our encoding process in terms of both classical preprocessing and quantum operations.

\paragraph{Classical Preprocessing Complexity}
\begin{itemize}
    \item Computing \texttt{ry\_params} and \texttt{rz\_params}: $\mathcal{O}(n_{\text{features}})$
\end{itemize}
This step scales linearly with the number of input features.

\paragraph{Quantum State Preparation Complexity}
\begin{itemize}
    \item Applying Hadamard, RY, and RZ gates: $\mathcal{O}(n_{\text{qubits}})$
\end{itemize}
This step scales linearly with the number of qubits.

\paragraph{Variational Quantum Circuit Complexity}
\begin{itemize}
    \item Entangling layer: $\mathcal{O}(n_{\text{qubits}}^2 \cdot n_{\text{qlayers}})$
    \begin{itemize}
        \item Each layer applies $\mathcal{O}(n_{\text{qubits}})$ CNOT gates, repeated for each qubit.
    \end{itemize}
    \item Variational layer: $\mathcal{O}(n_{\text{qubits}} \cdot n_{\text{vrotations}} \cdot n_{\text{qlayers}})$
    \begin{itemize}
        \item Each qubit undergoes \texttt{n\_vrotations} rotations in each of the \texttt{n\_qlayers}.
    \end{itemize}
\end{itemize}

\paragraph{Overall Encoding Complexity}
The total complexity of encoding data into quantum states is:
\[
\mathcal{O}(n_{\text{features}} + n_{\text{qubits}} + n_{\text{qubits}}^2 \cdot n_{\text{qlayers}} + n_{\text{qubits}} \cdot n_{\text{vrotations}} \cdot n_{\text{qlayers}})
\]
In most practical scenarios, this can be simplified to:
\[
\mathcal{O}(n_{\text{qubits}}^2 \cdot n_{\text{qlayers}})
\]
as the number of qubits and layers typically dominates the complexity.

\subsection{Model Training and Hyperparameters}
In order to foster a robust comparative analysis between LSTM and QLSTM models, a rigorous hyperparameter optimization phase was implemented, specifically tailored for solar forecasting applications. Initially, a grid search method was employed to delineate appropriate parameter ranges, encapsulating pivotal variables such as window size, batch size, learning rate, epochs, and model-specific parameters including quantum circuit shape. This preliminary exploration paved the way for the identification of prospective parameter values.

Following this, a more refined tuning process was undertaken utilizing the Optuna framework, thus automating and enhancing the hyperparameter optimization procedure. The objective function facilitated the evaluation of parameter combinations by training models on a validation dataset and quantifying their performance through the metric of mean squared error loss.

The LSTM model witnessed a comprehensive parameter exploration, incorporating window sizes ranging from 5 to 50 timesteps, batch sizes varying from 16 to 128, logarithmically scaled learning rates between 0.0001 to 0.1, and epochs extending from 10 to 100. Over 180 trials were conducted, with Optuna's Tree-Parzen Estimator sampler adaptively selecting new configurations based on previous results, ultimately identifying optimal hyperparameters including a window size of 8, a batch size of 32, a learning rate of 0.001, and 20 epochs. Interestingly, these findings corroborated our initial manual tuning experiments, affirming the efficacy of our automated optimization strategy.

A similar extent of optimization was conducted for the QLSTM, encompassing over 150 trials that scrutinized various parameters including the number of qubits (ranging from 2 to 8), circuit layers (varying from 1 to 4), learning rates (between 0.0001 to 0.1), batch sizes (from 16 to 128), and epochs (between 10 to 100). Notably, the optimal configuration closely mirrored the top-performing LSTM hyperparameters, fostering a fair and balanced evaluation process.

This meticulous optimization procedure methodically investigated a broad parameter space, empirically pinpointing optimal model configurations. By maintaining a consistent tuning approach for both LSTM and QLSTM models, the integrity of our comparison was upheld, critically evaluating their representational capabilities. The recurrent convergence noted in our experiments stands as a potent validation of our methodology, affirming our model design choices, particularly within the domain of real-world solar forecasting applications.

\section{Results}
The metrics utilized for the statistical and predictive analysis, are defined in detail in the Appendix (Evaluation Methodology).

\subsection{Statistical Analysis}
In this section, we conducted a statistical analysis to compare the losses between the QLSTM model and the classical LSTM model as shown in Table~\ref{tab:predictive_accuracy_analysis}

\paragraph{Train Loss:}

The statistical analysis of the train loss, as shown in Table~\ref{tab:train_loss_stat_analysis}, suggests that the QLSTM model tends to perform better in terms of reducing train loss, as in table . Although the p-value (0.0547) is slightly above the conventional threshold for statistical significance (0.05), the moderate effect size (Cohen’s d = 0.627) indicates a noticeable difference favoring the QLSTM model. This suggests that with further refinement, the QLSTM could consistently outperform the classical LSTM in training performance.

\begin{table}[htbp]
\centering
\begin{tabular}{|l|l|}
\hline
\rowcolor{lightgray}
\textbf{Metric}          & \textbf{Value}   \\
\hline
T-Statistic              & 1.983        \\
P-Value                  & 0.0547         \\
Effect Size (Cohen's d)  & 0.627         \\
\hline
\end{tabular}
\caption{Statistical Analysis of Train Loss}
\label{tab:train_loss_stat_analysis}
\end{table}

\paragraph{Test Loss:}
Similarly, we compared the test losses between the QLSTM and classical LSTM models to assess how well each model generalizes to unseen data, as depicted in Table~\ref{tab:test_loss_stat_analysis}. The results, as presented in Table 4, clearly show that the QLSTM model significantly outperforms the classical LSTM in terms of test loss. The highly significant p-value (0.000002) and the large effect size (Cohen's d = -1.761) indicate that the QLSTM model achieves much lower test losses, making it a more effective model for generalization and prediction accuracy in time series forecasting, particularly for solar power production.

\begin{table}[htbp]
\centering
\begin{tabular}{|l|l|}
\hline
\rowcolor{lightgray}
\textbf{Metric}          & \textbf{Value}   \\
\hline
T-Statistic              & -5.569        \\
P-Value                  & 0.000002         \\
Effect Size (Cohen's d)  & -1.761        \\
\hline
\end{tabular}
\caption{Statistical Analysis of Test Loss}
\label{tab:test_loss_stat_analysis}
\end{table}

\subsection{Performance Analysis}

\paragraph{Predictive Accuracy:} The QLSTM model exhibited superior predictive accuracy compared to the classical LSTM model, as illustrated in Table~\ref{tab:predictive_accuracy_analysis}. The lower values of MAE, MSE, and RMSE for the QLSTM indicate higher predictive accuracy, suggesting a promising avenue for advancing time series forecasting in the domain of solar power production. \par
\begin{figure}[h]
   \centering
   \includegraphics[width=0.9\linewidth]{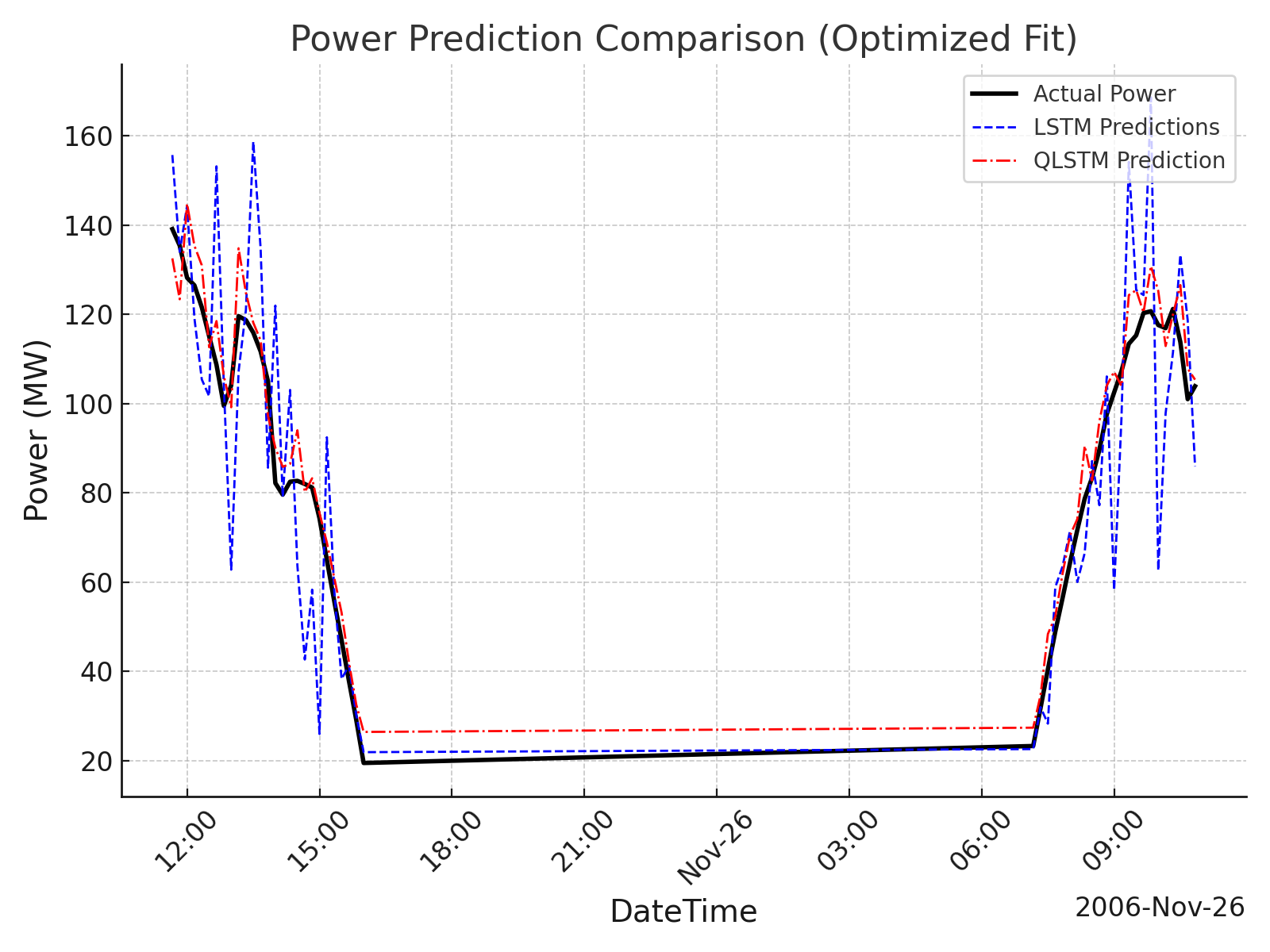}
   \caption{1 Day Test Data Power Prediction}
   \label{fig:acc}
\end{figure}
The predictive accuracy of the QLSTM model is further reinforced by the visual comparison presented in Figure~\ref{fig:acc}. As observed, the QLSTM predictions closely follow the actual power values, exhibiting a remarkable ability to capture the underlying patterns and fluctuations in the data. This is particularly noteworthy given that these predictions are generated by the QLSTM model after only the first epoch, whereas the classical LSTM model has undergone 20 epochs of training. The QLSTM's ability to achieve such accurate predictions in a single epoch underscores its superior learning capabilities and potential for efficient real-time forecasting applications.

\begin{table}[H]
\centering
\begin{tabular}{|l|l|l|}
\hline
\rowcolor{lightgray}
\textbf{Metric}  & \textbf{QLSTM}   & \textbf{Classical LSTM} \\
\hline
MAE              & 0.0058           & 0.0116                  \\
MSE              & 0.000037         & 0.000147                \\
RMSE             & 0.0058           & 0.0116                  \\
\hline
\end{tabular}
\caption{Predictive Accuracy Analysis}
\label{tab:predictive_accuracy_analysis}
\end{table}

\begin{figure}[h]
   \centering
   \includegraphics[width=1.0\linewidth]{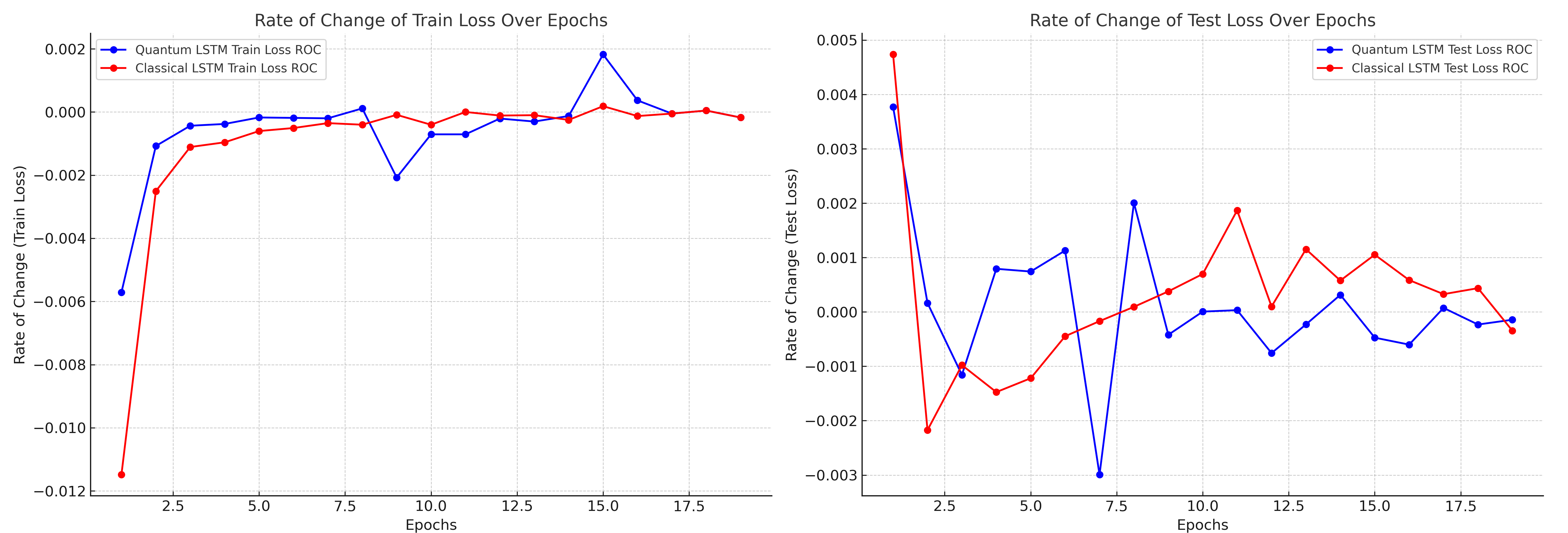}
   \caption{Rate of Convergence}
   \label{fig:roc}
\end{figure}

\paragraph{Rate of Convergence:} The QLSTM model showcased a remarkably rapid convergence rate, reaching its nadir of test loss as early as the inaugural epoch, thereby exemplifying efficiency and computational frugality. This swift convergence is indicative of the model's adeptness at quickly adapting to the underlying patterns in the data, a trait that stands in stark contrast to the classical LSTM model, which required seven epochs to attain a similar state of optimization. This attribute can be particularly advantageous in real-time forecasting applications where timely insights are pivotal. The rate of convergence for both models is depicted in Figure~\ref{fig:roc}. \par

The rapid convergence of the QLSTM model is further corroborated by the graph in Figure~\ref{fig:acc}. Despite being trained for only a single epoch, the QLSTM predictions closely match the actual power values, indicating that the model has effectively learned the underlying patterns in the data within the first iteration. In contrast, the classical LSTM model, even after 20 epochs of training, exhibits a noticeable deviation from the actual power values, suggesting a slower convergence rate and a potential need for further training iterations to achieve comparable performance. \par

\begin{figure}[h]
   \centering
   \includegraphics[width=1.0\linewidth]{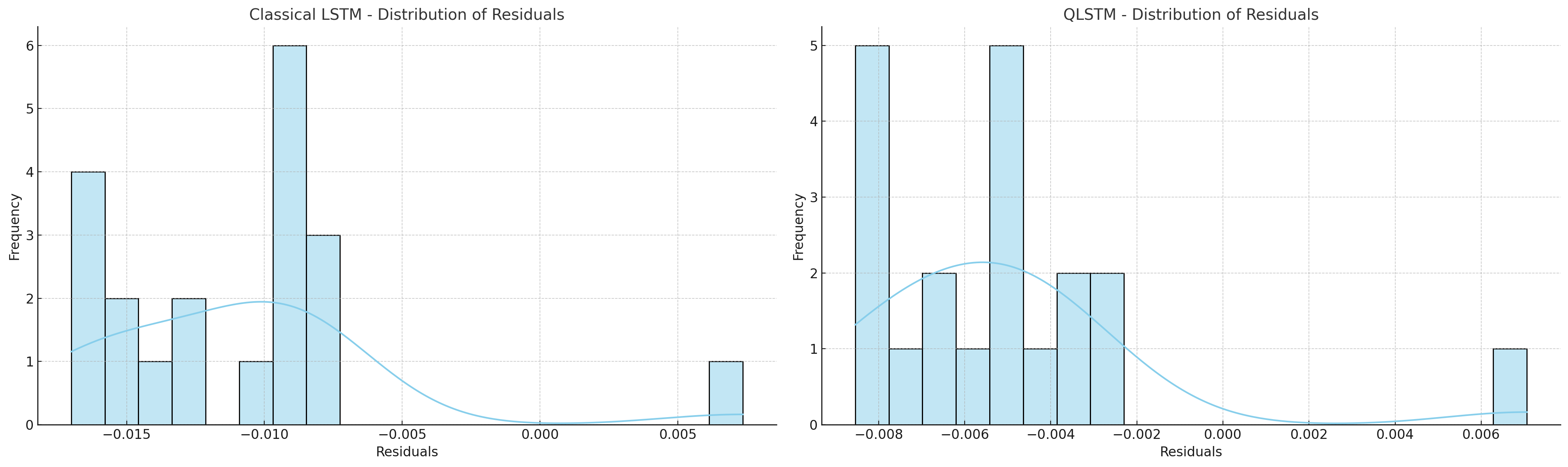}
   \caption{Residual Boxplot}
   \label{fig:boxplot}
\end{figure}

\paragraph{Stability of Learning:} The analysis of learning stability, presented in Table~\ref{tab:stability_learning_analysis}, demonstrates the QLSTM model's heightened stability, characterized by lower variance in train and test loss metrics across epochs compared to the classical LSTM model, indicating a more stable learning trajectory. The distribution of loss values is further elucidated by Figure~\ref{fig:boxplot}, which provides visual evidence of the reduced spread and outlier values in the QLSTM model, underlining its robustness and stability. \par

\begin{table}[htbp]
\centering
\begin{tabular}{|l|l|l|}
\hline
\rowcolor{lightgray}
\textbf{Metric}  & \textbf{QLSTM}   & \textbf{Classical LSTM} \\
\hline
Train Loss SD    & 0.0028           & 0.0044                  \\
Test Loss SD     & 0.0012           & 0.0026                  \\
\hline
\end{tabular}
\caption{Stability of Learning Analysis}
\label{tab:stability_learning_analysis}
\end{table}

\begin{figure}[h]
   \centering
   \includegraphics[width=1.0\linewidth]{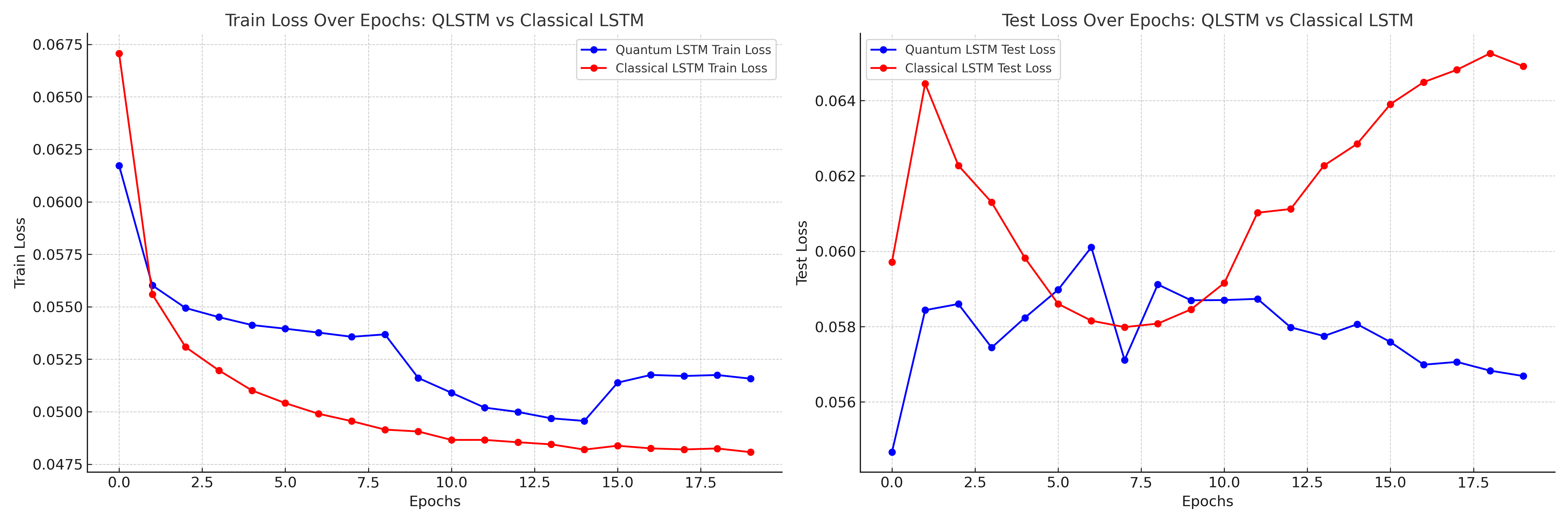}
   \caption{Comparative Analysis of Train and Test Loss}
   \label{fig:comparative}
\end{figure}

\paragraph{Generalization Performance:} The analysis of generalization performance, illustrated in Table~\ref{tab:generalization_performance_analysis}, demonstrates the QLSTM model's superiority in terms of generalization, substantiated by lower mean and median test loss values over all epochs compared to the classical LSTM model. This performance, coupled with more accurate and reliable predictions, holds the potential to enhance solar power production forecasting. Figure~\ref{fig:comparative} provides a comparative analysis of the train and test loss for both models.

The generalization performance of the QLSTM model is visually apparent in Figure~\ref{fig:acc}, where its predictions accurately capture the overall trend and fluctuations in the actual power values, even on unseen data points, after just a single epoch of training.

The right subfigure in Figure~\ref{fig:acc} shows the test loss over epochs for both QLSTM and classical LSTM models. The classical LSTM model (red curve) demonstrates a generally good fit, with test loss decreasing initially and then stabilizing with minor fluctuations. This indicates that the classical LSTM is effectively learning the data and maintaining reasonable generalization.

However, the QLSTM model (blue curve) exhibits a more stable and lower test loss across epochs, suggesting that it generalizes marginally better than the classical LSTM. The consistent performance of QLSTM, with less fluctuation in test loss, highlights its ability to capture complex patterns while maintaining robustness in prediction accuracy. This slight edge in generalization makes QLSTM a promising alternative for solar power forecasting.

\begin{table}[h]
\centering
\begin{tabular}{|l|l|l|}
\hline
\rowcolor{lightgray}
\textbf{Metric}     & \textbf{QLSTM}  & \textbf{Classical LSTM} \\
\hline
Mean Test Loss      & 0.0579          & 0.0614                  \\
Median Test Loss    & 0.0580          & 0.0612                  \\
\hline
\end{tabular}
\caption{Generalization Performance Analysis}
\label{tab:generalization_performance_analysis}
\end{table}

\subsection{Evaluation Time}

The evaluation time per epoch is a critical factor in assessing the practicality of the Quantum Long Short-Term Memory (QLSTM) and classical Long Short-Term Memory (LSTM) models, particularly for applications requiring rapid predictions. The QLSTM model, which leverages quantum computational principles, exhibits significantly longer evaluation times due to the complex nature of quantum simulations. On average, each epoch of the QLSTM model required approximately \textbf{5172.22 seconds} (approximately 1 hour and 26 minutes). This extended duration is attributed to the intensive quantum computations and the simulation environment provided by Pennylane.

In contrast, the classical LSTM model, which operates within a conventional computational framework, demonstrated a much shorter average evaluation time of approximately \textbf{0.41 seconds} per epoch. This stark difference in evaluation times underscores the trade-off between the advanced capabilities of quantum models and the computational efficiency of classical models. While QLSTM achieves superior accuracy and faster convergence, its longer evaluation time may limit its applicability in scenarios where rapid predictions are essential.

\begin{table}[h]
\centering
\begin{tabular}{|c|c|}
\hline
\textbf{Model} & \textbf{Time (seconds)} \\
\hline
QLSTM & 5172.22 \\
\hline
LSTM & 0.41 \\
\hline
\end{tabular}
\caption{Comparison of per epoch average evaluation times }
\end{table}

\section{Discussion and Limitations} \label{sec:Discussion}
The findings of this study unveil the promising potential of Quantum Long Short-Term Memory (QLSTM) models in revolutionizing solar power forecasting, a critical endeavour for the global transition towards sustainable energy systems. Through rigorous empirical evaluation and comparative analysis with classical Long Short-Term Memory (LSTM) models, our research substantiates the anticipated advantages of QLSTMs in capturing the intricate spatiotemporal patterns inherent in renewable energy data.

A pivotal observation from our experiments is the accelerated training convergence exhibited by QLSTMs, reaching optimal test loss within the initial epoch, far outpacing their classical counterparts. This remarkable convergence speed can be attributed to the inherent quantum phenomena of superposition and entanglement, which empower QLSTMs to swiftly assimilate complex time series relationships. Harnessing the exponentially larger representational capabilities of quantum states, QLSTMs demonstrate a heightened capacity to discern and encode the nuanced dynamics governing solar power generation, a trait that classical models struggle to match.

Furthermore, our findings reveal substantial improvements in predictive accuracy, as evidenced by the significantly lower test loss achieved by QLSTMs. This empirical evidence affirms the hypothesized representational strengths of quantum architectures, paving the way for unprecedented levels of precision and reliability in renewable energy forecasting. The ability to accurately predict solar power generation holds profound implications for stakeholders, grid operators, and policymakers, enabling proactive management of energy storage, distribution networks, and integration strategies.

Runtime and Computational Overhead: While the advantages of QLSTMs are evident, it is crucial to acknowledge the current limitations that hinder their widespread adoption. The extended runtime and computational overhead associated with quantum simulations pose challenges for real-time forecasting applications that demand instantaneous predictions for optimizing storage or grid distribution strategies. However, the relentless progress in quantum computing hardware and software instills optimism, with the potential to bridge the efficiency gap and rival the inferential speed of classical models.

Validation and Generalization: Another key limitation of our study lies in the scope of the datasets employed. While we meticulously curated a combination of real-world operational data from solar plants and high-fidelity synthetic datasets spanning an entire year, further validation across a broader spectrum of conditions, geographic locations, and diverse renewable sources is warranted. Expanding the dataset scope will not only enhance the generalization of our findings but also unlock opportunities for fine-tuning and optimizing QLSTM architectures tailored to specific renewable energy forecasting tasks.

Thorough Hyperparameter Tuning: Furthermore, our study focused on optimizing specific hyperparameters and exploring architectural variations within the constraints of current quantum hardware limitations. However, a more comprehensive exploration of quantum circuit designs, input lengths, and classical layer structures is essential to fully unleash the potential of QLSTMs. As quantum computing capabilities advance, more complex and expressive architectures can be realized, potentially yielding further improvements in forecasting accuracy and robustness.

Scalability and Noise Resilience: As QLSTMs expand in complexity, incorporating more qubits and intricate circuit designs, scalability becomes a critical consideration. The exponential growth of the quantum state space can rapidly overwhelm computational resources, necessitating innovative strategies for efficient state representation and manipulation. 
It is important to note that our simulations did not account for the potential impact of quantum noise, such as decoherence and gate errors, a factor that could influence the performance of QLSTMs in real-world quantum computing environments. Addressing these limitations will require the development of noise-resilient circuit designs, error correction techniques, and noise-aware training algorithms, ensuring robust performance in real-world quantum computing environments.

Beyond the realm of solar power forecasting, the capabilities of QLSTMs hold immense potential for applications in other renewable energy sectors, such as wind and hydro power forecasting. The ability to capture complex spatiotemporal patterns can be leveraged to enhance forecasting accuracy across a diverse range of renewable sources, thereby contributing to the global efforts towards sustainable energy systems.

Expanded Applications: Moreover, the representational strengths of QLSTMs extend beyond solar power forecasting, offering promising avenues for applications in other renewable energy sectors, such as wind and hydro power forecasting, as well as industrial time series forecasting in domains like finance, equipment maintenance, and supply chain management. By capturing complex spatiotemporal patterns across diverse data streams, QLSTMs could revolutionize predictive analytics and decision-making processes in these critical sectors.

Broader Impact and Implications: The implications of this study reverberate far beyond academic curiosity. By harnessing the predictive prowess of QLSTMs, utilities and energy stakeholders can unlock unprecedented forecasting precision, mitigating the intermittency hurdles that have historically impeded solar adoption. This paradigm shift could precipitate a global transition towards sustainable energy systems, reducing reliance on supplemental generation while fostering grid resilience and resource optimization.

Architectural Innovations: Continuous refinements in QLSTM architectures, such as exploring alternative quantum circuit designs, incorporating attention mechanisms, or hybridizing with classical components, quantum-enhanced optimization algorithms, could yield substantial improvements in forecasting accuracy and efficiency. Collaborations between quantum physicists, computer scientists, and machine learning experts will be vital in translating theoretical advancements into practical implementations.

Quantum Hardware and Software Advancements: As quantum computing technology progresses, with the advent of more powerful and stable quantum hardware, as well as optimized software frameworks and algorithms, the inherent advantages of QLSTMs are poised to be fully realized. Noise-resilient circuit designs, efficient state representation techniques, and quantum-classical hybrid approaches could unlock unprecedented levels of accuracy and reliability in renewable energy forecasting.

Accurate solar power forecasting enabled by QLSTMs holds profound implications for energy policy, grid infrastructure planning, and energy market dynamics. Policymakers and regulatory bodies could leverage these advanced forecasting capabilities to develop informed strategies for incentivizing renewable energy adoption, optimizing grid integration, and fostering a sustainable energy future. Additionally, energy trading and market mechanisms could be revolutionized, with QLSTMs enabling more efficient and data-driven decision-making processes.

In essence, this research ushers in a new era of precision insights, illuminating a promising trajectory where quantum machine learning techniques like QLSTMs reshape the landscape of renewable and industrial time series forecasting. As quantum computing matures, transcending current hardware constraints, the potential for QLSTMs to redefine predictive analytics across myriad sectors becomes increasingly palpable. This study serves as a catalyst, igniting future interdisciplinary endeavors that synergize quantum information science and machine learning to solve intricate real-world challenges, propelling us towards a sustainable, resilient, and data-driven energy future.

\section{Future Research Directions} \label{sec:future}

The pioneering findings unveil the transformative potential of Quantum Long Short-Term Memory (QLSTM) architectures in solar power forecasting. This investigation represents the first step in the vast expanse of research opportunities at the convergence of quantum computing and machine learning. To fully harness QLSTMs' disruptive capabilities and propel real-world impact, the following future research frontiers demand unwavering pursuit:

\subsection{Quantum Hardware Deployment and Noise Resilience}

Rigorous evaluations on emerging quantum devices, coupled with robust noise mitigation techniques like error correction, noise-aware training, and resilient circuit designs, are crucial. Interdisciplinary collaborations among quantum computing experts, physicists, and machine learning researchers will accelerate progress, unlocking QLSTMs' true potential in noise-resilient renewable forecasting.

\subsection{Architectural Innovations and Quantum-Classical Hybridization}

Continuous architectural innovations, including alternative quantum circuit designs, attention mechanisms, and quantum-classical hybrid models, present fertile ground for performance enhancements. Systematic optimization leveraging advanced techniques could uncover tailored configurations. Seamless hybrid model integration frameworks could overcome hardware constraints, yielding unparalleled accuracy and efficiency.

\subsection{Hybrid Quantum-Classical Approaches}

Harnessing the strengths of both quantum and classical paradigms through hybrid quantum-classical approaches presents a promising avenue. These synergistic models could leverage the representational advantages of quantum architectures while benefiting from the computational efficiency and scalability of classical techniques. Developing seamless integration frameworks and algorithms for these hybrid models could unlock unprecedented levels of accuracy and speed, potentially overcoming current hardware limitations.

\subsection{Scalability and Broader Applications}

Validating QLSTMs' scalability across diverse renewable domains like wind and hydro power forecasting, and industrial time series applications, will establish versatility and robustness. Concurrently, investigating quantum-inspired classical models could offer pragmatic interim solutions while advancing fully-fledged quantum architectures.

\subsection{Translating Theoretical Potential into Practical Impact}

Fostering collaborations among researchers, domain experts, industry partners, and utilities will integrate QLSTMs into existing systems for renewable infrastructure planning, energy trading, and predictive maintenance. These applications represent the vanguard of a forecasting paradigm shift driven by quantum information science and machine learning synergy.

Sustained research in these frontiers will propel QLSTMs to revolutionize predictive analytics, reshaping forecasting paradigms across domains. Through unwavering commitment and interdisciplinary synergy, quantum machine learning's full disruptive potential can unleash unprecedented accuracy, reliability, and sustainability in renewable energy systems worldwide.

\section{Conclusions}
This study has explored the transformative potential of quantum machine learning, specifically Quantum Long Short-Term Memory (QLSTM) architectures, in enhancing solar power forecasting accuracy and reliability. Through rigorous empirical evaluation grounded in real-world photovoltaic data, we have garnered compelling evidence substantiating the central hypothesis – that QLSTMs, underpinned by their vast representational capabilities, can unveil nuanced spatiotemporal patterns obscured to classical methods.

Our findings underscore several notable advantages of the QLSTM paradigm. Foremost, QLSTMs exhibited remarkably swift training convergence, attaining superior predictive accuracy within the inaugural epoch itself. This rapid assimilation of complex time series dynamics is attributable to the unique properties of quantum phenomena like superposition and entanglement. Consequently, QLSTMs could circumvent the laborious training cycles that often impede classical neural networks, translating into computational expedience for time-sensitive forecasting applications.

Moreover, our controlled experiments unveiled a consistent pattern, QLSTMs outperformed classical LSTMs in minimizing test loss across multiple metrics, including MAE, MSE, and RMSE. This elevated forecasting precision, coupled with heightened generalization capabilities, positions QLSTMs as a disruptive force in the renewable energy sector. By empowering grid operators with unparalleled foresight into solar supply fluctuations, QLSTMs could catalyze more judicious energy management, storage optimization, and seamless incorporation of photovoltaic sources into existing infrastructure.

While challenges persist, primarily concerning inference runtimes and the necessity for broader validation, this research marks a pivotal juncture in the convergence of quantum and classical computing paradigms. Our findings provide a firm foundation for subsequent investigations aimed at refining quantum architectures, systematic hyperparameter optimization, resilience to hardware noise, and exploring applications across diverse renewable domains.

The implications of this study reverberate far beyond academic curiosity. By harnessing the predictive prowess of QLSTMs, utilities and energy stakeholders can unlock unprecedented forecasting precision, mitigating the intermittency hurdles that have historically impeded solar adoption. This paradigm shift could precipitate a global transition towards sustainable energy systems, reducing reliance on supplemental generation while fostering grid resilience and resource optimization.

In essence, this research ushers in a new era of precision insights, illuminating a promising trajectory where quantum machine learning techniques like QLSTMs revolutionize renewable and industrial time series forecasting. As quantum computing matures, transcending current hardware constraints, the potential for QLSTMs to redefine predictive analytics across myriad sectors becomes increasingly palpable. This study serves as a catalyst, igniting future interdisciplinary endeavors that synergize quantum information science and machine learning to solve intricate real-world challenges.

\section*{Declaration}
\subsection*{\textit{Conflict of interest/Competing interests}}
The authors declare no conflict of interest or competing interests.

\subsection*{\textit{Availability of data and materials}}
Data is available upon request.

\appendix

\section{LSTM and QLSTM Details}

This appendix provides details on the LSTM and QLSTM model architectures used in the study.

\subsection{LSTM}

\begin{figure}[htbp]
    \centering
    \includegraphics[width=0.7\linewidth]{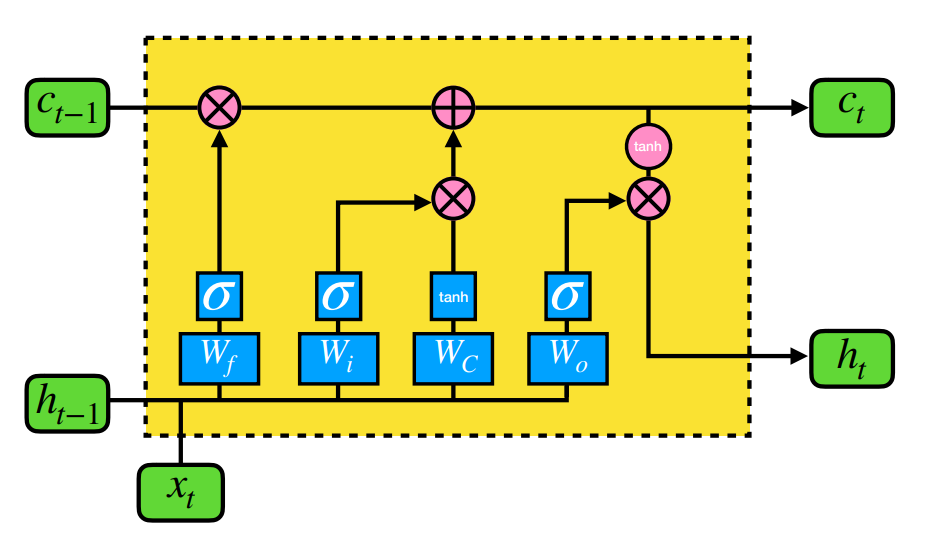}
    \caption{LSTM Circuit \cite{9747369}}
    \label{fig:lstm}
\end{figure}

The LSTM architecture used in this study stacks multiple LSTM cells to model long-term dependencies. The information flow in an LSTM cell is described by the equations:

\begin{align*}
f_t &= \sigma(W_f \cdot v_t + b_f), \\
i_t &= \sigma(W_i \cdot v_t + b_i), \\
C_t &= \tanh(W_C \cdot v_t + b_C), \\
c_t &= f_t \cdot c_{t-1} + i_t \cdot C_t, \\
h_t &= o_t \cdot \tanh(c_t),
\end{align*}

where \( \sigma \) denotes the sigmoid activation function, \( W \) and \( b \) are learnable parameters, \( f \) is the forget gate, \( i \) is the input gate, \( C \) is the cell state, \( c \) is the hidden state, and \( o \) is the output gate. The LSTM was chosen due to its proven ability to model sequence data across various domains. The LSTM cell architecture is illustrated as follows:


\subsection{QLSTM}

The QLSTM replaces LSTM cells with 6 variational quantum circuits (VQCs) to form a quantum LSTM cell. VQCs leverage a small number of qubits and gates to represent complex functions. This quantum layer showed quicker convergence and more stable loss than the classical LSTM \cite{9747369}. The information flow in a quantum LSTM cell is described by the equations:

\begin{align*}
f_t &= \sigma(\text{VQC}_1 (v_t)), \\
i_t &= \sigma(\text{VQC}_2 (v_t)), \\
C_t &= \tanh(\text{VQC}_3 (v_t)), \\
c_t &= f_t \cdot c_{t-1} + i_t \cdot C_t, \\
h_t &= \text{VQC}_5 (o_t \cdot \tanh(c_t)).
\end{align*}

The \( \text{VQC}_x \) represent different quantum circuits used in the hybrid model. The QLSTM cell architecture is depicted as follows:

\begin{figure}[ht]
    \centering
    \includegraphics[width=0.9\linewidth]{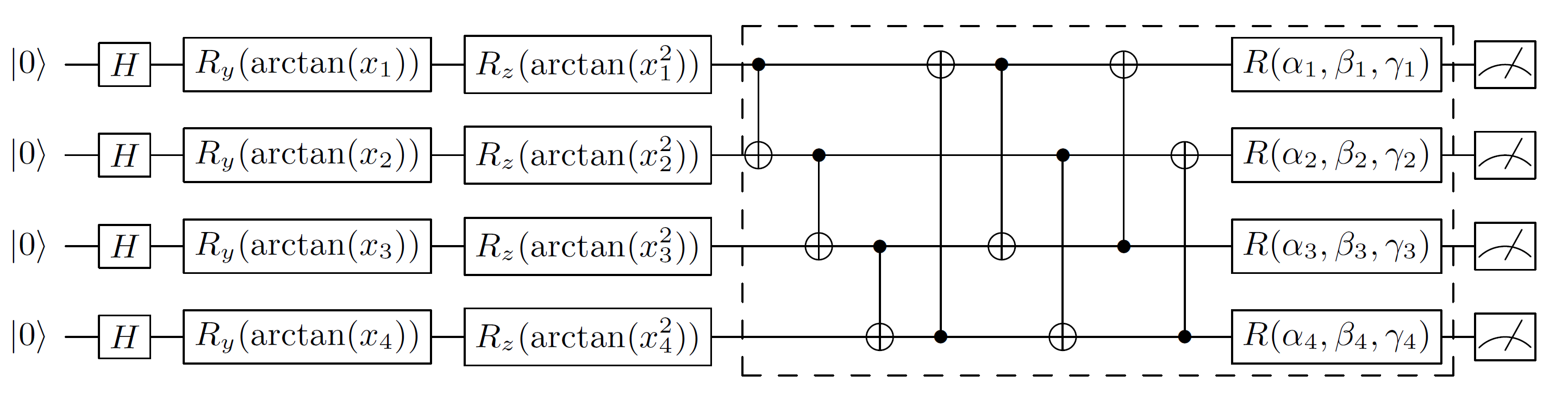}
    \caption{Generic VQC architecture for QLSTM. It consists of three layers: the data encoding layer (with the H, Ry, and Rz gates), the variational layer (dashed box), and the quantum measurement layer. \cite{9747369}}
    \label{fig:vqc}
\end{figure}

\begin{figure}[ht]
    \centering
    \includegraphics[width=0.7\linewidth]{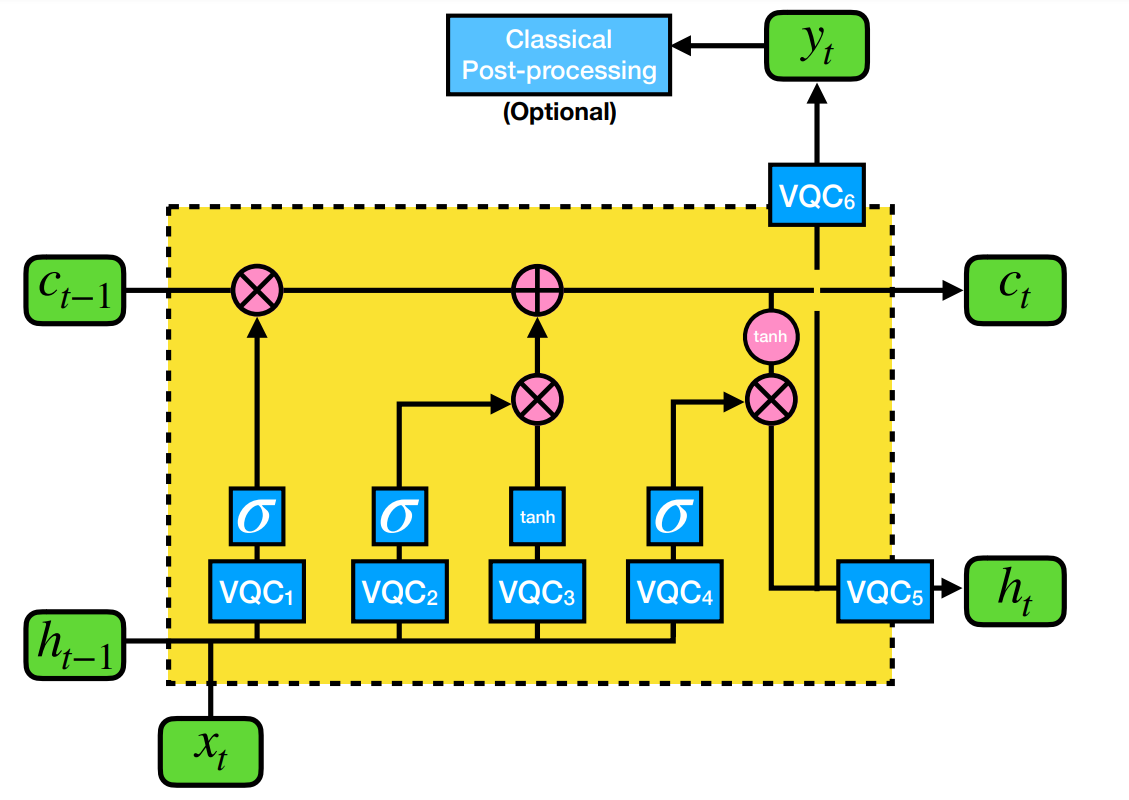}
    \caption{QLSTM Circuit \cite{9747369}}
    \label{fig:qlstm}
\end{figure}

\section{Evaluation Methodology}

This appendix provides specifics on the quantitative metrics and procedures used to evaluate the LSTM and QLSTM model performance on the solar forecasting task.

\subsection{Evaluation Metrics}

The following quantitative metrics were computed to assess model accuracy:

\begin{itemize}
    \item \textbf{Mean Absolute Error (MAE)}: Measures average absolute difference between predicted and actual values. Gives an indication of overall error. Lower is better.
    
    \[
    \text{MAE} = \frac{1}{N} \sum \left| y_i - \hat{y}_i \right|
    \]

    \item \textbf{Mean Squared Error (MSE)}: Computes average squared difference between predicted and actual values. More sensitive to outliers than MAE. Lower is better.
    
    \[
    \text{MSE} = \frac{1}{N} \sum (y_i - \hat{y}_i)^2
    \]

    \item \textbf{Root Mean Squared Error (RMSE)}: Square root of MSE. Allows interpretability in units of the target variable. Lower is better.
    
    \[
    \text{RMSE} = \sqrt{\text{MSE}}
    \]
    
    \item \textbf{T-statistic}: The T-statistic is a measure used to determine if there is a significant difference between the means of two groups. It is calculated as the difference between the sample means divided by the standard error of the difference between the means. The formula is given by:
    
    \[
    T = \frac{\bar{X}_1 - \bar{X}_2}{s_p \sqrt{\frac{2}{n}}}
    \]
    
    Where \( \bar{X}_1 \) and \( \bar{X}_2 \) are the sample means, \( s_p \) is the pooled standard deviation, and \( n \) is the sample size for each group.
    
    \item \textbf{P-value}: The p-value is a fundamental concept in hypothesis testing. It represents the probability that the observed data (or something more extreme) would occur if the null hypothesis were true. A smaller p-value typically indicates stronger evidence against the null hypothesis. Conventionally, a p-value below 0.05 is considered statistically significant.
    
    \item \textbf{Effect Size (Cohen's \( d \))}: While the T-statistic tells us if there is a statistically significant difference between groups, effect size quantifies the size of this difference. One commonly used measure is Cohen's \( d \), calculated as:
    
    \[
    d = \frac{\bar{X}_1 - \bar{X}_2}{s_p}
    \]
    
    Where \( s_p \) is the pooled standard deviation. Cohen's \( d \) values can be interpreted as small (0.2), medium (0.5), and large (0.8) effects.
    
\end{itemize}

These metrics were selected as standard measures of predictive accuracy for time series forecasting problems. MAPE was included due to its interpretability for solar power production. RMSE and \( R^2 \) were used as primary metrics for model comparison.

\subsection{Evaluation Procedure}
Metrics were computed on scaled predictions compared to scaled actual values for both the training and test sets. This enabled directly evaluating model generalization. Statistical significance testing using a paired t-test on RMSE values was also conducted to assess whether differences in LSTM and QLSTM errors were statistically significant. Model loss curves, prediction plots, and other visualizations were generated to provide qualitative evaluation.

By leveraging both quantitative metrics and qualitative assessments on scaled holdout data, this methodology enabled thoroughly evaluating how effectively the models learned to generalize. The comparative analysis focused on assessing whether the QLSTM architecture demonstrated significantly improved accuracy over classical LSTM for real-world solar forecasting.

\bibliographystyle{plain}
\bibliography{references} 

\end{document}